# An important discovery-—with no electric field, the center of the electron cloud of K atom does not coincide with the nucleus

Pei-Lin You

Institute of Quantum Electronics- Guangdong Ocean University, Zhanjiang 524025, China.

It is a general point of view that with no electric field, the nucleus of an atom is at the center of the electron cloud, so that all kinds of atoms do not have permanent electric dipole moment (EDM). In fact, the idea is untested. Using two special capacitors containing Potassium vapor we discovered that the electric susceptibility $x_e$ of K atoms is directly proportional to the density N, and inversely to the absolute temperature T, as is polar molecule. We have distinguished between permanent and induced dipole moments carefully. There is good evidence that a ground state neutral K atom has a large permanent EDM, $d_K$=[1.58±0.19(stat)±0.13(syst)] × $10^{-8}$e.cm. A direct proof of time-reversal violation in nature occurred in K atoms. The experimental K material is supplied by Strem Chemicals Co. USA. If K atom has a large permanent EDM, why the linear Stark effect has not been observed? The article discussed the question thoroughly. These experimental results can be repeated in other laboratories.

## 1. Introduction

The realization that there is one small nucleus, which contains the entire positive charge and almost the entire mass of the atom, is due to the investigations of Rutherford, who utilized the scattering of alpha particles by matter[1]. Since the time of Rutherford it was commonly believed that in the absence of an external field, the nucleus of an atom is at the centre of the electron cloud, so that any kind of atoms does not have permanent electric dipole moment (EDM)[1-4]. Therefore, there is no polar atom in nature except for polar molecules. In fact, the idea is untested. We hope to test the idea by precise measurement. The experimental K material with purity 99.95％ is supplied by Strem Chemicals Co.USA(see Fig.7). Our experiments showed that with no electric field, the center of the electron cloud of K atom does not coincide with its nucleus, and ground state neutral K atom has a large EDM. This result is the product of past ten years of intense research [13-21].

In order for an atom or elementary particle to possess a EDM, time reversal (T) symmetry must be violated, and through the CPT theorem CP(charge conjugation and parity) must be violated as well[2]. In fact, despite the relentless search for a non-zero EDM of atom or elementary particle for more 50 years, no conclusive results have been obtained so far[2-4]. The non-zero observation of EDM in any non-degenerate system will be a direct proof of time-reversal violation in nature[2-4]. On the other hand, some evidence for CP violation beyond the Standard Model comes from Cosmology. Astronomical observations indicate that our universe is mostly made of matter and contains almost no anti-matter. The first example of CP violation was discovered in 1964, but it has been observed only in the decays of the $K_o$ mesons. After 38 years, the BaBar experiment at Stanford Linear Accelerator Center (SLAC) and the Belle collaboration at the KEK laboratory in Japan announced the second example of CP violation. "The results gave clear evidence for CP violation in B mesons. However, the degree of CP violation now confirmed is not enough on its own to account for the matter-antimatter imbalance in the universe." (SLAC Press Release July 23, 2002). Theorists found it hard to see why CP symmetry should be broken at all and even harder to understand why any imperfection should be so small[2]. "Such rare decays are very challenging to identify, but over the past few years we have learned to cleanly isolate one event in a million." (SLAC Press Release Sep 28, 2006). This fact suggests that there must be other ways in which CP symmetry breaks down and more subtle effects must be examined. So EDM experiments are now considered an ideal probe for evidence of new sources of CP violation. If an EDM is found, it will be compelling evidence for the existence of new sources of CP violation. Our three experiments showed that the ground state neutral K atom is polar atom and new example of CP violation occurred in K atoms. It is a classic example of how understanding of our universe advances through atomic physics research. Our experimental apparatus is still kept, and we welcome anyone who is interested in the experiments to visit and examine it. Because the EDM of an atom is extremely small, we applied several ingenious experimental techniques [14]. Although the three experiments are simple, no one has completed the experiments using K atoms so far! Few experiments in atomic physics have produced a result as surprising as this one.

R.P. Feynman considered a polar molecule which carries a permanent dipole moment $d_o$, such as water molecule [5]. With no electric field, the individual dipoles point in random directions, and no macroscopic dipole moment is observed. But when an electric field is applied, the individual dipoles tend to orient in the direction of the field. On the other hand, when atoms are placed in an electric field, they become polarized, acquiring *induced* electric dipole moments in the direction of the field. Note that the electric susceptibility($x_e$) caused by the orientation of polar molecules is inversely proportional to the absolute temperature(T): $x_e = B/T$ while the induced electric susceptibility due to the distortion of electronic motion in atoms or molecules is temperature independent: $x_e = A$, where A and the slope B is constant. The electric susceptibility $x_e = C/C_o - 1$, $C_o$ is the vacuum capacitance and C is the capacitance of the capacitor filled with the material. This difference in temperature dependence offers a means of separating the polar and non-polar substances experimentally[6].

The electric susceptibility of a gaseous polar dielectric is[6]

$$x_e = A + N d_o^2 / 3kT \varepsilon_o = A + B/T \qquad (1)$$

where k is Boltzmann constant, $\varepsilon_o$ is the permittivity of free space, N is the number density of molecules and A is induced electric susceptibility. R.P. Feynman checked Eq.(1) with the orientation polarization experiment of water vapor. He plotted the straight line from four experimental points [5]. The following table 1 gives the four experimental data [7].

| Table 1 | T(K) | Pressure(cm Hg) | $x_e$ |
|---|---|---|---|
| | 393 | 56.49 | $400.2 \times 10^{-5}$ |
| | 423 | 60.93 | $371.7 \times 10^{-5}$ |
| | 453 | 65.34 | $348.8 \times 10^{-5}$ |
| | 483 | 69.75 | $328.7 \times 10^{-5}$ |

From the ideal gas law, the number density of water vapor is $N = P/kT = 1.392 \times 10^{25}$ m$^{-3}$. From the four experimental data, we work out the induced electric susceptibility $A = 17.8 \times 10^{-5} \approx 0$. The slope is according to (1)

$$B = N d_o^2 / 3k \varepsilon_o \qquad (2)$$

We work out the slope $B = N d_o^2 / 3k \varepsilon_o = 1.50K$ and the EDM of H$_2$O molecule is $d_{H2O} = (3k\varepsilon_o B /N)^{1/2} \approx 6.28 \times 10^{-30}$ C·m, while the observed value is $d_{H2O} = 6.20 \times 10^{-30}$ C·m [6].

By measuring capacitance at different temperatures, it is possible to distinguish between permanent and induced dipole moments[6]. Obviously the electric susceptibility caused by the orientation of polar molecules is directly proportional to the density N, and inversely to the absolute temperature T. If K atom is the polar atom, the same form $x_e \propto N/T$ should be expected in measuring capacitance. This article reported three interesting experimental curves of the electric susceptibility of K atoms in external electric field, which have not been observed in the history of physics!

## 2. Experimental method and result

The first experiment: investigation of the relationship between the electric susceptibility of K vapor and its number density N. The experimental apparatus is a closed glass container. It resembles a Dewar flask in shape. The external and internal diameters of the container are $D_1=80.8$mm and $D_2=56.8$mm. The external and internal surfaces of the container are plated with silver, respectively shown by **a** and **b** in Fig.1. These two silver layers constitute the cylindrical capacitor. The length of the two silver layers is L=24cm and the total length is L'=33cm. The thickness of the glass wall is $\triangle=1.5$mm. The width of the gap that will be filled by K vapor is $H_1=9.0$mm. The capacitance of the cylindrical capacitor is measured by a digital meter. The precision of the meter was 0.1pF, the accuracy was 0.5% and the surveying voltage was V=1.2 volt. This capacitor is connected in series by two capacitors. One is called C' and contains the K vapor of thickness $H_1$, another one is called C'' and contains the glass medium of thickness $2\triangle$. The total capacitance C is

$$C = C'C''/ (C'+C'') \quad \text{or} \quad C' = C'' C / (C'' - C) \qquad (3)$$

where C'' and C can be directly measured. The experiment to measure C'' is easy. We can made a cylindrical capacitor of glass with thickness $2\triangle$ and put it in a temperature-control stove. By measuring capacitances at different temperatures, we can find C''. When the container is empty, it is pumped to a vacuum pressure $P \leq 10^{-8}$ Pa in the range 30℃~300℃ for 20 hours. The aim of the operation is to remove impurities such as oxygen adsorbed on the inner walls of the container. We measured the total capacitance C = 49.4 pF and another one C''=1658 pF, the vacuum capacitance is $C'_0 =51.9$ pF according to (3). The mass of a potassium sample is 5g. The next step, the K sample with purity 0.9995 and supply by Strem chemicals company

USA(see Fig.7), is put in the container. The container is again pumped to vacuum pressure P $\leq 10^{-8}$ Pa at room temperature, then it is sealed. We obtain the experimental apparatus as shown in Fig.1. Now, the measured capacitance of the cylindrical capacitor is no longer 51.9 pF but it is C'=378 pF at room temperature, where C=307.8 pF and C''=1658 pF. Since the K material enters into the glass container, the capacitance has increased obviously. We put the capacitor into a temperature-control stove, raise the temperature of the stove very slow and the susceptibility of K vapor is measured at different temperature in the range 30℃ to 260℃. The following table 2 gives the experimental data.

**Table 2    The susceptibility of K vapor at different temperature (C'$_0$ =51.9 pF)**

| T (℃) | 30 | 36 | 42 | 48 | 54 | 60 | 66 | 72 | 78 | 84 | 90 | 96 | 102 | 108 |
|---|---|---|---|---|---|---|---|---|---|---|---|---|---|---|
| C'(pF) | 379 | 381 | 382 | 383 | 385 | 388 | 389 | 390 | 390 | 391 | 391 | 392 | 393 | 394 |
| $x_e$ | 6.30 | 6.34 | 6.36 | 6.38 | 6.42 | 6.48 | 6.50 | 6.51 | 6.51 | 6.53 | 6.53 | 6.55 | 6.55 | 6.59 |
| T (℃) | 114 | 120 | 126 | 132 | 138 | 144 | 150 | 156 | 162 | 168 | 174 | 180 | 186 | 192 |
| C'(pF) | 395 | 396 | 397 | 398 | 399 | 400 | 403 | 406 | 410 | 414 | 420 | 425 | 432 | 440 |
| $x_e$ | 6.61 | 6.63 | 6.65 | 6.67 | 6.69 | 6.70 | 6.76 | 6.82 | 6.90 | 6.98 | 7.09 | 7.19 | 7.32 | 7.48 |
| T (℃) | 198 | 206 | 208 | 214 | 220 | 226 | 232 | 238 | 244 | 248 | 252 | 256 | 258 | 260 |
| C'(pF) | 446 | 455 | 451 | 430 | 408 | 366 | 334 | 302 | 278 | 268 | 264 | 258 | 256 | 254 |
| $x_e$ | 7.59 | 7.77 | 7.69 | 7.28 | 6.86 | 6.05 | 5.43 | 4.82 | 4.36 | 4.16 | 4.09 | 3.97 | 3.93 | 3.90 |

Now we keep the temperature of the stove at $T_1$ =260℃=533K for 4 hours. The susceptibility of K vapor is measured at different time when the temperature keeps constant $T_1$. The table 3 gives the experimental data.

**Table 3 The susceptibility of K vapor at different time when the temperature keeps at $T_1$ =533(K))**

| t (min) | 0 | 5 | 10 | 15 | 20 | 25 | 30 | 35 | 40 | 45 | 50 | 60 |
|---|---|---|---|---|---|---|---|---|---|---|---|---|
| C'(pF) | 254 | 441 | 664 | 882 | 1079 | 1344 | 1547 | 1775 | 1977 | 2206 | 2465 | 2865 |
| $x_e$ | 3.90 | 7.50 | 11.8 | 16.0 | 19.8 | 24.9 | 28.8 | 33.2 | 37.1 | 41.5 | 46.5 | 54.2 |
| t (min) | 70 | 80 | 90 | 100 | 120 | 140 | 160 | 180 | 190 | 210 | 230 | 240 |
| C'(pF) | 2978 | 3084 | 3171 | 3254 | 3326 | 3383 | 3420 | 3448 | 3456 | 3456 | 3456 | 3456 |
| $x_e$ | 56.4 | 58.4 | 60.1 | 61.7 | 63.1 | 64.2 | 64.9 | 65.4 | 65.6 | 65.6 | 65.6 | 65.6 |

The experiment showed that with the prolongation of isothermal time, both the number density and the susceptibility of K vapor simultaneously increases rapidly. When the isothermal time t≥190 minutes at $T_1$ =260℃, we measured the total capacitance $C_t$ = 2270 pF and the capacitance of the glass medium $C_t$''=6615 pF, the capacitance of the K vapor is $C_t$' =3456 pF according to (3). So the susceptibility of K vapor remains constant $x_e = C_t'/C'_o$ − 1=65.6. It means that the reading of the susceptibility is obtained under the condition of K saturated vapor pressure, both the number density and the susceptibility appear to be stable. The formula of saturated vapor pressure of K atoms is P= 10 $^{7.183-4434.3/T}$ psi (1 psi =6894.8Pa) [8]. The effective range of the formula is 533K ≤T ≤1033K . Using the formula we obtain the saturated vapor pressure of K vapor $P_k$=503.5Pa at 533K. From the ideal gas law, the number density of K vapor $N_1 = P_K /kT_1$ =6.84×10$^{22}$ m$^{-3}$. The experimental curve is shown in Fig.2. Note that when the isothermal time t≤60 minutes , the plot is a straight line. Because the density N of K vapor is proportional to the isothermal time t, the straight line shows that the electric susceptibility $x_e$ of K vapor is directly proportional to its density N when T keeps constant.

It is well known that the electric susceptibility is of the order of 10$^{-3}$ for any kind of gases, for example 0.0046 for HCl gas, 0.007 for water vapor [6]. Note that $x_e$ = C' / C'$_0$ − 1=65.6>>1 for K vapor, the experimental result exceeded all expert's expectation!

The second experiment: investigation of the relationship between $x_e$ of K vapor and T at a fixed density. The apparatus was a closed glass container but the K vapor was at a fixed density $N_2$. In order to control the quantity of K vapor, the container is connected to another small container that contains K material by a glass tube from the top. We put these two containers into a temperature-control stove are slowly heated. To keep the temperature of the stove at T =230℃ for 3 hours and the designed experimental container is sealed. Thus we obtain a cylindrical capacitor filled with K vapor at a fixed density. The apparatus shown in Fig.3 and two stainless steel tubes **a** and **b** build up the cylindrical capacitor. The external and internal diameters of the two tubes are $D_3$=67.2mm and $D_4$=52.2mm. The plate separation is $H_3$=7.5mm and the plate area is $S_3$=4.00×10$^{-2}$ m$^2$. The capacitance C was still measured by the digital meter and the vacuum capacitance $C_{20}$=47.2pF. By measuring electric susceptibility of K vapor at different temperature, we obtain $x_e$ =A+B/T≈B/T, where the intercept A≈0 and the slope of the line B≈190(K). The experimental results are shown in Fig.4. The orientation polarization of K atoms has been observed, as polar molecules $H_2O$ (where B≈1.50 K )[5].

The third experiment: measuring the capacitance of K vapor at various voltages (V) under a fixed density $N_2$ and a fixed temperature $T_2$. The apparatus was the same as the preceding one and $C_{30}$ = $C_{20}$ = 47.2 pF. The measuring method is shown in Fig. 5. C was the capacitor filled with K vapor to be measured and kept at $T_2$=303K. $C_d$ =520pF was used as a standard capacitor. Two signals $V_c(t)=V_{co} \cos \omega t$ and $V_s(t)=V_{so} \cos \omega t$ were measured by a two channel digital real-time oscilloscope (supply by Tektronix TDS 210 USA). The two signals had the same frequency and always the same phase at different voltages. From Fig.5, we have $(V_s-V_c)/V_c = C/C_d$ and $C=(V_{so}/V_{co} - 1)C_d$. In the experiment $V_{so}$ can be adjusted from zero to 800V. The measured result of the capacitance at different voltages is shown in Fig.6. When $V_1=V_{co}$≤0.4volt, $C_1$=232pF or $x_e$=3.92 is constant. With the increase of voltage, the capacitance decreases gradually. When $V_2=V_{co}$=400 volt, $C_2$=53.0pF or $x_e$=0.1229, it approaches saturation. If all the dipoles in a gas turn toward the direction of the field, this effect is called the saturation polarization. The $x_e$-V curve showed that the saturation polarization of the K vapor is approaching when E≥$V_2/H_2$=5.4×10$^4$V/m.

**3. Theory and interpretation** The local field acting on a molecule in a gas is almost the same as the external field **E**[6]. The electric susceptibility of a gaseous polar dielectric is[9]

$$x_e = NG + N d_o L(a)/ \varepsilon_o E \qquad (4)$$

where a = $d_o$ E /kT, $d_o$ is EDM of a molecule, NG =A and G is the molecule polarizability. The Langevin function is

$$L(a) = [(e^a + e^{-a})/(e^a - e^{-a})] - 1/a \qquad (5)$$

The Langevin function L(a) is equal to the mean value of cos θ ( θ is the angle between **d$_o$** and **E**) [9]:

$$<\cos \theta> = \mu \int_0^p \cos \theta \exp(d_o E \cos \theta /kT) \sin \theta \, d\theta = L(a), \quad \mu = [\int_0^p \exp(d_o E \cos \theta /kT) \sin \theta \, d\theta]^{-1} \qquad (6)$$

where μ is a normalized constant. This result shows that L(a) is the percentage of polar molecule lined up with the field in the total number. When a<<1 and L(a)≈a/3, when a>>1 and L(a)≈1[9].

The next step, we will consider how this equation is applied to K atoms. Due to the atomic polarizability of K atoms is G= 43.4×10$^{-30}$ m$^{3[10]}$, the number density of K atoms N≤10$^{23}$ m$^{-3}$ and the induced susceptibility A=NG≤4.34×10$^{-6}$ can be neglected. In addition, the induced dipole moment of K atoms is $d_{int}$ =G $\varepsilon_o$E[10], due to E≤10$^5$v/m in the experiment, then $d_{int}$≤3.9×10$^{-35}$ C.m can be neglected. From Eq.(4) we obtain

$$x_e = Nd L(a)/\varepsilon_o E \qquad (7)$$

where d is the EDM of an K atom and N is the number density of K vapor. L(a)= <cos θ> is the percentage of K atoms lined up with the field in the total number. Note that E=V/H and $\varepsilon_o$= $C_o$ H / S, leading to

$$C - C_o = \beta L(a)/a, \qquad (8)$$

where β = S N d$^2$/kTH is a constant. **This is the polarization equation of K atoms**. Due to a=dE/kT= dV/kTH we obtain the formula of atomic EDM

$$\mathbf{d_{atom} =(C - C_o )V / L(a)SN} \qquad (9)$$

In order to work out **L(a)** and **a** of the first experiment, note that in the third experiment when the field is weak ($V_1$=0.4V), $a_1$ <<1 and L($a_1$)≈$a_1$/3. From Eq.(8): $C_1 - C_{30}$=232 - 47.2= β /3 and β =554.4pF. When the field is strong($V_2$=400V), $a_2$ >>1 and $C_2 - C_{30}$ = L($a_2$) β /$a_2$. We work out $a_2$ = 94.6, L($a_2$)=L(94.6)=0.9894. Due to a=dV/kTH, so a/$a_2$ =V$T_2H_2$ /$T_1H_1V_2$, Substituting the corresponding values, **a** = 0.134 and L(a)=0.0447. L(a)=0.0447 shows that only 4.47％ of K atoms are lined up with the direction of the field in the first experiment.

Notice that we deduced Eq(8) and Eq (9) from the formula of the parallel-plate capacitor $\varepsilon_o$= $C_o$ H / S, so the cylindrical capacitor must be regarded as a equivalent parallel-plate capacitor with the plate area $S_1$= $C'_o H_1/ \varepsilon_o$. Substituting the values: $S_1$= 5.28×10$^{-2}$ m$^2$, $N_1$=6.84×10$^{22}$ m$^{-3}$, V=1.2volt and C - $C_o$ = C'$_t$ - C'$_0$ = 3404.1pF, from Eq. (9) we work out

$$d_K = (C - C_o)V / L(a) S_1 N_1 = 2.53 \times 10^{-29} \text{C.m} = 1.58 \times 10^{-8} \text{e.cm} \qquad (10)$$

The statistical error about the measured value is $\triangle d_1/d \leqslant 0.12$, considering all sources of systematic error is $\triangle d_2/d \leqslant 0.08$, and the combination error is $\triangle d/d \leqslant 0.15$. We find that

$$d_K = [2.53 \pm 0.30(\text{stat}) \pm 0.20(\text{syst})] \times 10^{-29} \text{C.m} = [1.58 \pm 0.19(\text{stat}) \pm 0.13(\text{syst})] \times 10^{-8} \text{e. cm} \qquad (11)$$

**Although above calculation is simple, but no physicist completed the calculation up to now!**

**4. Discussion**

①The formula $d_{atom} = (C - C_o)V/L(a)SN$ can be justified easily. The dipole moment of an K atom is $d = e \cdot r$. N is the number of K atoms per unit volume. L(a) is the percentage of K atoms lined up with the field in the total number. When an electric field is applied, the K atoms tend to orient in the direction of the field as dipoles. On the one hand, the change of the charge of the capacitor is $\triangle Q = (C - C_0)V$. On the other hand, due to the volume of the capacitor is SH, the total number of K atoms lined up with the field is SHNL(a). The number of layers of K atoms which lined up with the filed is H/r. Because inside the K vapor the positive and negative charges cancel out each other, the polarization only gives rise to a net positive charge on one side of the capacitor and a net negative charge on the opposite side. Then $\triangle Q = SHN L(a)e / (H/r) = SN L(a)e r = SN L(a)d = (C - C_0)V$, so the EDM of an K atom is $d = (C - C_0)V / SN L(a)$.

②If K atom has a large EDM, why the linear Stark effect has not been observed? As a concrete example, let us treat the linear Stark shifts of the hydrogen( n=2) and K atom. Notice that the fine structure of the hydrogen (n=2) is only 0.33 cm$^{-1}$ for the Hα lines of the Balmer series, where $l = 656.3$ nm, and the splitting is only $\triangle l = 0.33 \times (656.3 \times 10^{-7})^2 = 0.014$ nm, therefore the fine structure is difficult to observe [1]. The linear Stark shifts of the energy levels is proportional to the field strength: $\triangle W = d_H E = 1.59 \times 10^{-8}$ E e.cm. When $E = 10^5$ V/cm, $\triangle W = 1.59 \times 10^{-3}$ eV, this corresponds to a wavenumber of 12.8 cm$^{-1}$. So the linear Stark shifts is $\triangle l = \triangle W l^2/hc = 12.8 \times (656.3 \times 10^{-7})^2 = 0.55$ nm. It is so large, in fact, that the Stark shift of the lines of the hydrogen is easily observed [1]. However, the most field strength for K atoms is $E_{max} = 5.4 \times 10^4$ V/m, if K atom has a large EDM $d_K = 1.58 \times 10^{-8}$ e.cm, and the most splitting of the energy levels of K atoms $\triangle W_{max} = d_K E_{max} = 8.53 \times 10^{-6}$ eV. This corresponds to a wavenumber of $6.87 \times 10^{-2}$ cm$^{-1}$. On the other hand, the observed values for a line pair of the first primary series of K atom (Z=19, n=4) are $l_1 = 769.90$ nm and $l_2 = 766.49$ nm[8]. The most linear Stark shift of K atoms is only $\triangle l = \triangle W (l_1 + l_2)^2 / 4hc = 0.0041$ nm. **It is so small, in fact, that a direct observation of the linear Stark shifts of K atom is not possible!**

③The shift in the energy levels of an atom in an electric field is known as the Stark effect. Normally the effect is quadratic in the field strength, but first excited state of the hydrogen atom exhibits an effect that is linear in the strength. This is due to the degeneracy of the excited state. This result shows that the hydrogen atom (the quantum number n=2 ) has very large EDM, $d_H = 3e a_o = 1.59 \times 10^{-8}$ e.cm ($a_o$ is Bohr radius) [11]. L.D. Landay once stated that "The presence of the linear effect means that, in the unperturbed state, the hydrogen atom has a dipole moment"[11]. The alkali atoms having only one valence electron in the outermost shell can be described as hydrogen-like atoms[12]. Since the quantum number of the ground state alkali atoms are n≥2 rather than n=1( this is 2 for Li, 3 for Na, 4 for K, 5 for Rb and 6 for Cs), as the excited state of the hydrogen atom. So we conjecture that the ground state neutral alkali atoms may have large EDM of the order of e $a_o$ [13].

④From 1ev=kT we get T=11594K, in the temperature range of the experiment 303K ≤T≤533K, kT<<1ev so the measured capacitance change ( C'$_t$ − C'$_0$ ) entirely comes from the contribution of ground state K atoms.

Accurate measurements of the EDM of Rubidium (Rb) and Cesium (Cs) atom in ground state have been carried out. Similar results have been obtained [15-19]. Because no any scientist has observed the saturation polarization effect of any gas till now, so the saturation polarization of K, Rb or Cs vapor in ordinary temperatures is an entirely unexpected discovery and another two articles discussed the question [20-21].

**Acnowledgement**

The author thank to Prof. Zhen-Hua Guo, Prof. Xiang-You Huang, Prof. Wei-Min Du, Dr. Yo-Sheng Zhang, Dr. Min-Jun Zheng, thank to our colleagues Rui-Hua Zhou, Zhao Tang, Xue-Ming Yi, Shao-Wei Peng, and Engineer Yi-Quan Zhan, Engineer Jia You for their help in the work.

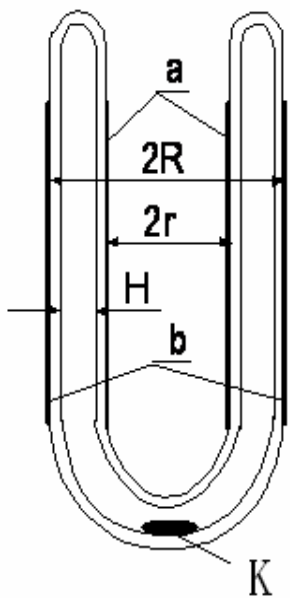

**Fig.1** This is the longitudinal section of the apparatus. The apparatus is a glass Dewar flask filled with K vapor. Silver layers **a** and **b** build up a cylindrical glass capacitor.

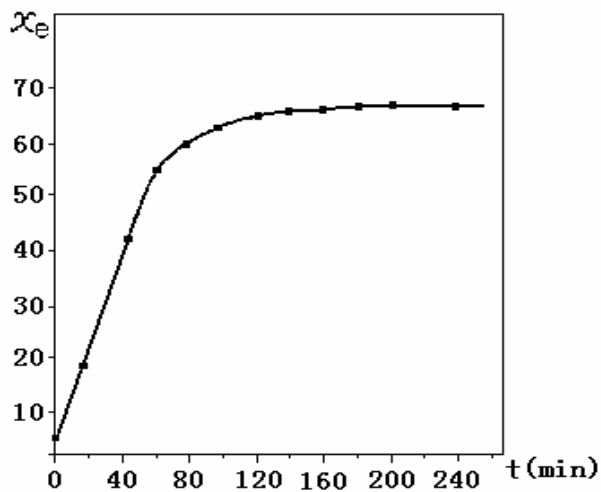

**Fig.2** The curve showed that. the electric susceptibility $x_e$ of K atoms varies in direct proportion to the density N, as polar molecules.

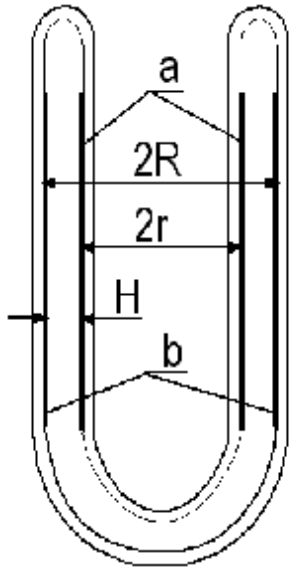

**Fig.3** The longitudinal section of the apparatus of another two experiments. The glass Dewar flask filled with K vapor at a fixed density. Two stainless steel tubes **a** and **b** build up the capacitor.

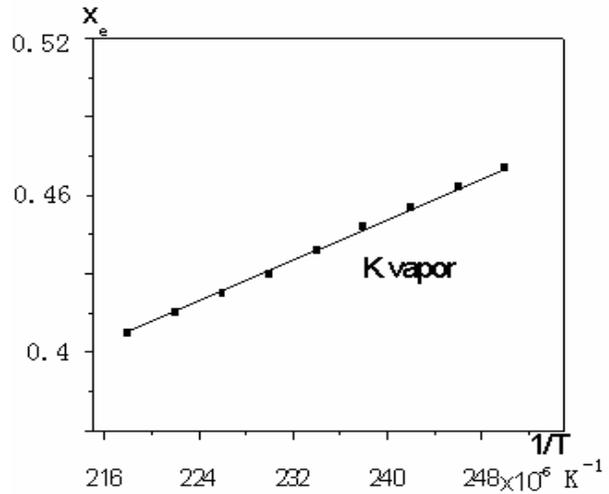

**Fig.4** The curve showed that the electric susceptibility $x_e$ of K atoms varies inversely proportional to the temperature T. The slope of the line is B≈190(k).

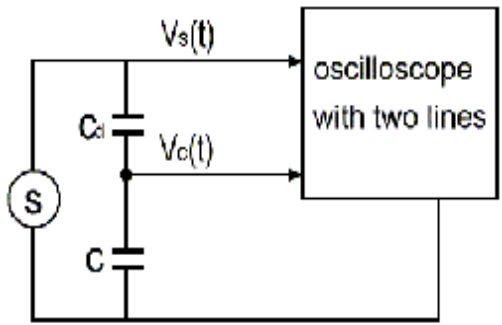

**Fig.5** The diagram shows the measuring method. C is capacitor filled with K vapor to be measured and Cd is a standard one, where $V_s(t) = V_{so} \cos \omega t$ and $V_c(t) = V_{co} \cos \omega t$.

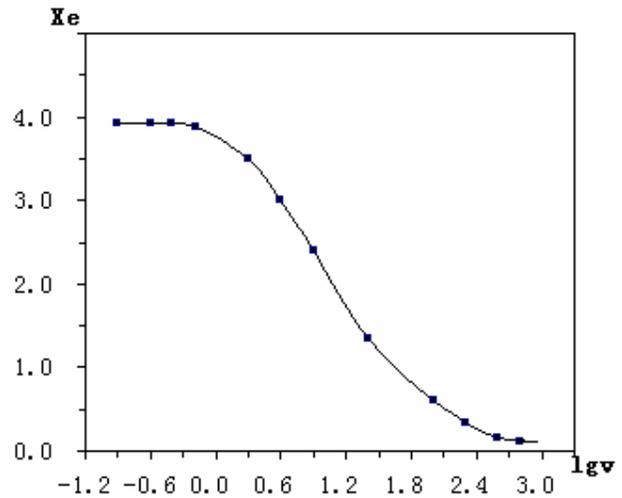

**Fig.6** The experimental curve shows that the saturation polarization effect of the K vapor is approached when E≥5.4×10$^4$v/m.

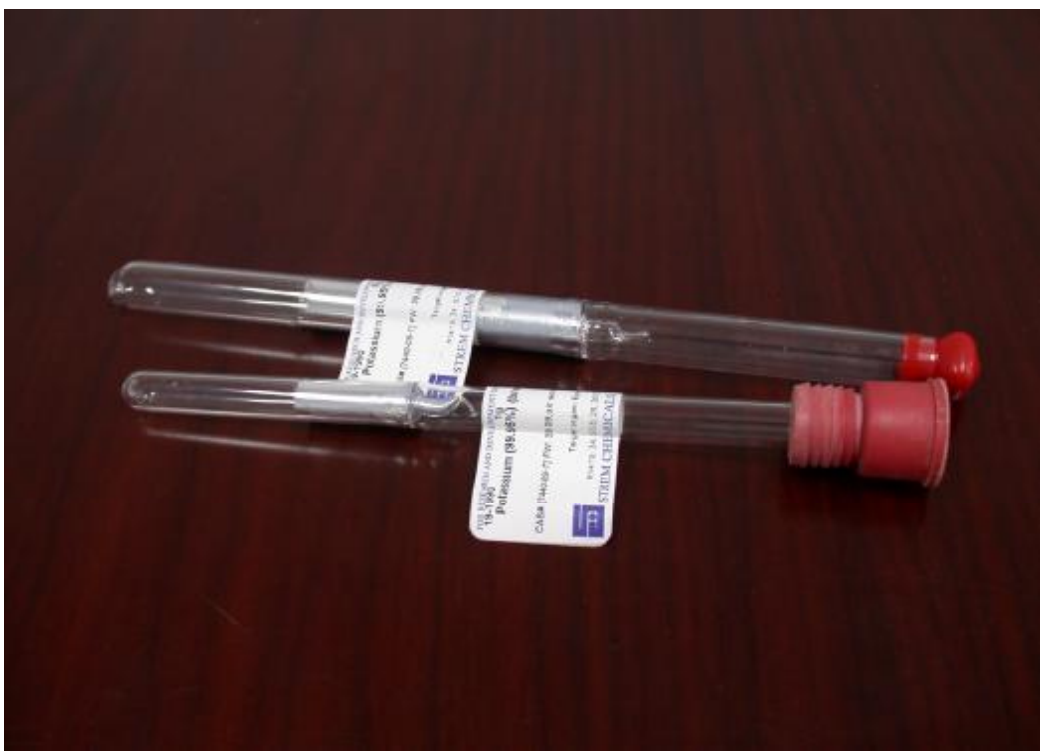

**Fig.7** The experimental potassium material with purity 99.95％ is supplied by STREM CHEMICALS Co. USA. The mass of the two potassium sample are 1g and 5g respectively. They are in their respective breakseal ampoules.